\documentclass[aps,pra,preprint,amsmath,amssymb]{revtex4}
\usepackage{graphicx,epsfig,epsf}
\usepackage{color}
\usepackage{amsfonts}
\usepackage{wasysym}
\usepackage{textcomp}
\usepackage[T1]{fontenc}
\usepackage{amssymb,amsmath, amsthm}

\def\ket#1{|#1\rangle}

\begin{document}
\sloppy
\title{Large effects of boundaries on spin amplification in spin chains} 
\author{Benoit Roubert$^{(1,2)}$, Peter Braun$^{(3,4)}$ and Daniel Braun$^{(1,2)}$}
\affiliation{{$^{(1)}$ Laboratoire de Physique Th\'eorique  -- IRSAMC, Universit\'e de Toulouse, UPS, F-31062 Toulouse, France}\\
{$^{(2)}$ LPT -- IRSAMC, CNRS, F-31062 Toulouse, France}\\
$^{(3)}$ Fachbereich Physik, Universit\"at Duisburg--Essen, 47048 Duisburg,
  GERMANY\\
  $^{(4)}$ Institute of Physics, Saint-Petersburg University,
198504 Saint-Petersburg,  RUSSIA} 
\begin{abstract}
We investigate the effect of boundary conditions on spin amplification in spin 
chains.  We show that the boundaries play a crucial
role for the dynamics: A single additional coupling between the first and
last spins can
macroscopically modify the physical behavior compared to the open chain,
even in the limit of infinitely long chains. We show that this effect can be
understood in terms of a ``bifurcation'' in Hilbert space that can give
access to different parts of Hilbert space with macroscopically different
physical properties of the basis functions, depending
on the boundary conditions.  On the technical side, we
introduce semiclassical methods whose precision increase with
increasing chain length and allow us to analytically demonstrate the
effects of the boundaries in the thermodynamic limit. 
\end{abstract}
\maketitle
\section{Introduction}
Quantum state transfer through spin chains has attracted considerable
attention starting with a seminal paper by Bose \cite{Bose03}. In such a
scheme, an initial quantum state is prepared on a spin at one end of a
chain, whereas all other spins are in a pre-defined state, say all pointing
down. The  
system of coupled spins is then let to evolve freely, and after a certain
time, long enough for a spin wave to propagate to the other end of the
chain, the quantum state of the last spin is read-out.  The last spin
becomes in 
general entangled with the rest of the chain, and one therefore obtains a
mixed state when ignoring the rest of the chain. 
Bose showed that the fidelity of such a transfer through an un-modulated
spin chain with fixed nearest-neighbors Heisenberg couplings exceeds the
maximum classically possible value for up to 80 spins.  This work has been
generalized in several directions \cite{Christandl05,De_chiara05,Karbach05,Yung05,Furman06,Kay06,Lyakhov07,Di_franco08,Wiesniak08}.  Substantial effort was
spent to increase the fidelity of the state 
transfer. Perfect state transfer was predicted for chains with couplings
that increase like a square root as 
function of position along the chain towards
the center of the chain, leading effectively to a rotation of a large
collective spin 
\cite{Christandl04}. Also, reducing the coupling between the terminating  
spins of the 
chain and the rest of the chain was shown to provide a recipe for perfect
state transfer, at the cost of slowing down the transfer \cite{Wojcik05}. It
was noted that arbitrarily high fidelity could 
also be achieved through dual-rail encoding \cite{Burgarth05a} in two chains,
even with randomly coupled chains \cite{Burgarth05}. \\
Spin chains have also been studied in the context of spin
amplification. Detecting single spins, and even more so, to measure their
state is a 
formidable challenge \cite{Rugar04}. Lee and Khitrin proposed a clever
scheme of a ``quantum domino'', where an initially flipped spin leads to the
propagation of a domain wall and ultimately the copying of the initial spin
state onto a GHZ like state,
$(\alpha\ket{\downarrow}+\beta\ket{\uparrow})\ket{\downarrow}^{\otimes(N-1)}\rightarrow\alpha\ket{\downarrow}^{\otimes
  N}+\beta\ket{\uparrow}^{\otimes N}$. Kay  noted a connection between
quantum state transfer and spin amplification, which allowed to map
insights from optimal state transfer to optimal spin amplification
\cite{Kay07} and vice versa. Indeed, the same representation of the Hamilton
operator in the two cases can be obtained by exchanging the couplings and
basis functions at the same time.  For spin amplification, one wants basis
functions with a single domain wall and a hamiltonian which flips just the
spin adjacent to the domain wall, inducing the domino effect.  For quantum
state transfer, the interesting basis functions all have a single excitation
located on one of the $N$ spins, and the hamiltonian consists of
nearest-neighbors exchange couplings.  

Recently, there has been interest in geometrical and topological effects in
spin-networks. Quantum state transfer was extended to more
complicated networks, in particular hypercubes \cite{Christandl04}, and to
quantum computing during the transfer \cite{Kay05}. It was  
shown that during the transfer along a chain, an arbitrary single
qubit rotation can be performed by appropriately splitting and recombining
the chain. Even two qubit gates can be performed by coupling incoming
and outgoing  
chains that carry the qubits to and from a central chain. Also, topological
quantum gates were proposed, in which the chain is 
closed to a ring, and threaded by an appropriate Aharonov-Bohm flux
\cite{Kay05}. Very 
recently, topological effects were exploited for locally controlling the
dynamics in spin-networks \cite{Burgarth09}.\\  

In this paper we study the influence of the boundaries on spin amplification
in spin chains.  One might think that the boundaries consisting of the two
terminating spins should play a negligible role in the limit of very long
chains, $N\to\infty$.  The effects of boundary conditions indeed
vanish in the thermodynamical limit in most situations in physics. 
Well--known exceptions  exist in the presence of long-range correlations, as
for example exactly at a quantum phase transition \cite{DBraun98b}.
But, surprisingly, it turns out that in spin chains, far away from any phase
transition, the boundary conditions
can 
drastically modify the dynamical behavior.  The presence or absence of a
single additional coupling between the last and the first spin can lead to
macroscopically different 
time-dependent polarization even in the limit of arbitrarily long chains. We
will demonstrate that a kind of ``bifurcation'' in Hilbert space can take
place that explains the macroscopically different
behavior.  Depending on the boundary conditions, basis functions can be
reached which may differ both in their physical properties, as well as in the
scaling of the dimension of the basis with $N$. Furthermore, we will show 
that there are different ways of closing a chain to a ring, which not only drastically 
modify the physical behavior of the chain, but even lead to nonequivalent matrix
representations of the hamiltonian, and different dynamics in the accessible
part of Hilbert space. There is even a way of closing the chain such that
the different topology is felt in one part of the Hilbert space, but not in
another. 

To study the chains in the limit of very large $N$, we introduce an
innovative semiclassical approach, the precision of which increases with
increasing $N$.  The method works well for dimensions of the relevant basis
which scale linearly with $N$, and allows us to prove persistent macroscopic
differences in the physical behavior for $N\to\infty$.  
We start the analysis by reviewing the ``quantum domino'' system introduced
in \cite{Lee_Khitrin_2005}, and developing the semiclassical method at the
example of
linear chains with open boundary conditions (simply called ``linear
chains'') in the following.  
\section{Linear chain}
\label{seq:Linear-chain}
\subsection{Description of the system}
\label{seq:Linear-chain-Description-of-system}
We consider a linear chain of $N_L$ spins $-\frac{1}{2}$ with nearest-neighbors interactions, whose hamiltonian is given by
\begin{eqnarray}
\label{eq:Linear-chain-Hamiltonian}
H^{(L)} = \frac{J}{2} \sum_{k=2}^{N_L-1}X_k \left( 1_{k-1} \otimes 1_{k+1} - Z_{k-1}\otimes Z_{k+1}\right).
\end{eqnarray}
$X_k$, $Z_k$ are Pauli operators acting on spin $k$. The coupling constant $J$ will be set to $J=1/2$ throughout the paper. $H^{(L)}$ is an effective hamiltonian derived in \cite{Lee_Khitrin_2005} for a one-dimensional Ising chain of two level atoms with nearest-neighbors interactions, irradiated by a weak resonant transverse field. The hamiltonian has the physical meaning that when a spin $S_k$ is surrounded by two spins ($S_{k-1}$ and $S_{k+1}$) of opposite sign, the operator $X_k$ flips spin $S_k$. In the original model, this is achieved by the dependence of the resonance frequency of an atom on the state of its neighbors. 
If we consider the situation where the system is initially in the state
$|\Lambda_1 \rangle=|\downarrow_1 \uparrow_2 \ldots
\uparrow_{N_L} \rangle$, i.e.~the first spin is down, all the others
up, the dynamic is restrained to evolve in a subset of the total Hilbert
space of size $N_L-1$. Initially $H^{(L)}$  couples $\left| \Lambda_1\right
\rangle$ to $\left| \Lambda_2 \right \rangle = \left|\downarrow_1
\downarrow_2\uparrow_3 \ldots \uparrow_{N_L} \right\rangle$. Then in general
the system couples  $\left| \Lambda_k \right \rangle= \left|\downarrow_1
\ldots\downarrow_k \uparrow_{k+1} \ldots \uparrow_{N_L} \right\rangle$ ($k$
spins down, all the others up with  $k=2,\ldots,N_L-2$) to $\left|
\Lambda_{k-1} \right \rangle$ and  $\left| \Lambda_{k+1} \right
\rangle$. Finally, at the end of the chain, $\left| \Lambda_{N_L-1} \right
\rangle$ is reached, which is itself only coupled back to $\left|
\Lambda_{N_L-2}\right \rangle$. On the other hand, the state where all spins
are initially up, is an eigenstate of $H^{(L)}$ of eigenvalue $0$ and is
therefore stationary. Thus, a stimulated wave of flipped spins can be
triggered by the flip of a single spin. In other words the system acts as a
spin amplifier that amplifies the initial states $\left| \downarrow\right
\rangle $ or $\left| \uparrow\right \rangle $ to a macroscopic polarization
of the entire chain. Note that the restriction to the small subspace of
$N_L-1$ states $\left| \Lambda_1 \right\rangle, \ldots, \left|
\Lambda_{N_L-1} \right\rangle $ is a consequence of the fact that the
initially excited spin is at the beginning of the chain. A single excited
spin in the middle of the chain leads to significantly different dynamics
(see section \ref{sec.cc}).  
The states $\left\{ \left| \Lambda_{k} \right\rangle \right\}$, $k \in \left[ 1,\ldots,N_L-1 \right]$, form an orthonormal basis in which $H^{(L)}$ is represented by
\begin{eqnarray}
\label{eq:Linear-chain-Matrix-representation-of-hamiltonian}
H^{(L)} =  \frac{1}{2}\left(
\begin{array}{ccccc}
   0 & 1      &\,      &\,   & \,  \\
   1 & \ddots & \ddots&  \,   &  \, \\
\,&\ddots &\ddots &\ddots  &\, \\
   \,&\, &\ddots &\ddots  &1 \\
   \,& \, &\, &1  &0 \\
\end{array}
\right) ~.
\end{eqnarray}
In the subspace considered, $H^{(L)}$ is therefore equivalent to a 1D tight-binding hamiltonian
with constant nearest-neighbors hopping, $\displaystyle H^{(L)} = \frac{1}{2} \sum_{k=1}^{N_L-2}
 \left |\Lambda_k \right \rangle  \left \langle \Lambda_{k+1} \right | + h.c.  $, whose eigenstates are Bloch waves,
\begin{eqnarray}
\left|\Phi_p^{(L)}\right\rangle = \sqrt{\frac{2}{N_L}}\sum_{k=1}^{N_L-1}\sin\left[ \frac{p \pi k }{N_L}\right]\left|\Lambda_k \right\rangle ~.
\end{eqnarray}
The corresponding eigenvalues form a 1D energy band, 
\begin{eqnarray}
\label{eq:Linear_Chain_Eigenvalues}
\lambda_p^{(L)} = \cos \left(\frac{p \, \pi}{N_L}\right), p=1,\ldots,N_L-1\,.
\end{eqnarray}
The knowledge of the exact eigenvalues and eigenvectors of $H^{(L)}$ allows us to obtain an analytical expression of the propagator,
\begin{eqnarray}
\label{eq:Linear-chain-Propagator}
U_{k,k_{0}}^{(L)}(t) &=&\left\langle \Lambda_k \right\vert
e^{-i H^{(L)} t}\left\vert\Lambda_{k_0}\right\rangle   \nonumber \\
&=&\sum_{p=1}^{N_L-1}\left\langle\Lambda_{k} \right\vert \left. \Phi^{(L)}
_{p}\right\rangle e^{-it\lambda _{p}^{(L)}}\left\langle \Phi^{(L)}_{p}\right\vert
\left. \Lambda_{k_0} \right\rangle  \nonumber \\
&=&\sum_{p=1}^{N_L-1}F^{(L)}_{k,k_{0}}(p \, , t) ~, \nonumber\\
F^{(L)}_{k,k_0}\left(p \, , t \right) &\equiv &\frac{2}{N_L} \sin\left(\frac{p \pi k }{N_L}\right)\sin\left(\frac{p \pi k_0 }{N_L}\right) e^{- it \cos \frac{p \pi}{N_L }} ~. 
\end{eqnarray}
In \cite{Lee_Khitrin_2005}, this form of the propagator was used to study
numerically the time dependent mean polarization. In spite of the
exponential simplification of the problem in the subspace considered,
compared to the dynamics in the full $2^{N_L}$ dimensional Hilbert space,
each of the $(N_L-1)^2$ matrix elements of the propagator
(\ref{eq:Linear-chain-Propagator}) still contains a sum of ${\cal O}(N_L)$
terms. We now show that very precise approximations of $U^{(L)}$ can be
obtained that only involve one or few terms. This allows us to obtain closed
analytical expressions for the time dependent mean
polarization. We propose two such approximations. The first one is based on
an exact representation of $U^{(L)}$ in terms of Bessel functions. The
second is of semiclassical nature. For a given $t$, both become the more
precise the larger $N_L$, i.e.~the longer the chain. 
\subsection{Representation of the propagator in terms of  Bessel functions }
\begin{figure}[ht]
\includegraphics[width=9cm]{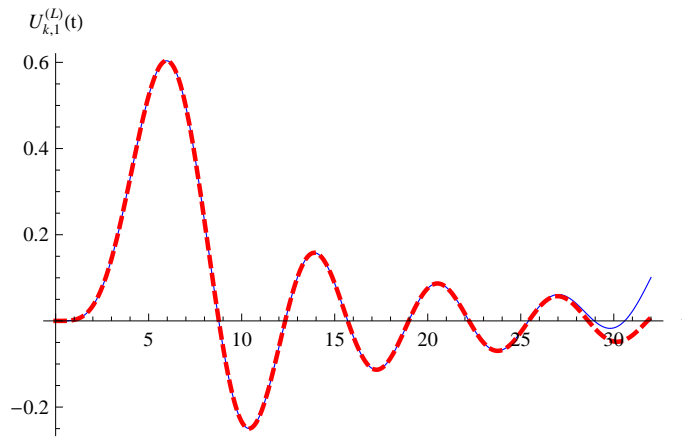}
\caption{(Color online)  Approximation with a single Bessel function ($m=0$ in Eq.(\ref{eq:Linear-chain-Propagator-in-terms-of-Bessel-functions}), dashed red line) of $U^{(L)}_{k,1}(t)$ (continuous blue line) for $k=5, N_L=20 $ }
\label{fig:Linear-chain-Comparison-Exact-Bessel-Approximated-Bessel-One-term}
\end{figure}

\begin{figure}[ht]
\includegraphics[width=9cm]{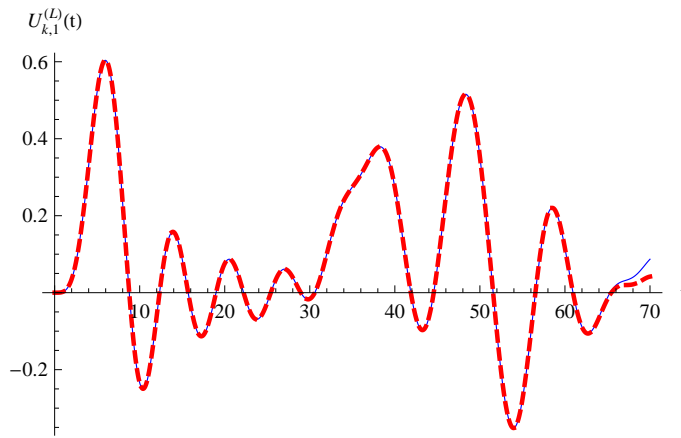} 
\caption{(Color online) Approximation with three Bessel functions ($m=-1,0,1$ in Eq.(\ref{eq:Linear-chain-Propagator-in-terms-of-Bessel-functions}), dashed red line) of $U^{(L)}_{k,1}(t)$ (continuous blue line) for $k=5, N_L=20 $}
\label{fig:linear-chain-Comparison-Exact-Bessel-Approximated-Bessel-Three-terms}
\end{figure}

Considering that $F^{(L)}_{k,k_{0}}(p \, , t)=F^{(L)}_{k,k_{0}}(-p \, , t)$, and that  $%
F^{(L)}_{k,k_{0}}(0\, ,t)=F^{(L)}_{k,k_{0}}(N_L\, ,t)=0$, we can double the summation range and evaluate the  sum by 
Poisson summation, 
\begin{eqnarray}
U^{(L)}_{k,k_{0}}(t) &=&\frac{1}{2}\sum_{p\, =-N_L}^{N_L-1}F^{(L)}_{k,k_{0}}(p\, , t) \nonumber \\
&=&\frac{1}{2}\sum_{m \, = -\infty }^{\infty }\int_{-N_L-\frac{1}{2}
}^{N_L-\frac{1}{2}}e^{i 2\pi m p}F^{(L)}_{k,k_{0}}(p\, ,t) ~ dp ~ .
\end{eqnarray}
We have $F^{(L)}_{k,k_{0}}(p\, ,t)=F^{(L)}_{k,k_{0}}(2N_L+p\, ,t)$, i.e.~all matrix elements have a  $2 N_L$ periodicity in $p$. Hence, the integrals are over a period of the integrand and we are allowed to shift the integration interval as we like. Setting $x=\pi p/N_L$ we arrive at the propagator in terms of the Bessel functions with the argument $t,$
\begin{eqnarray*}
U^{(L)}_{k,k_{0}}(t)&=&\frac{1}{2}\sum_{m=-\infty }^{\infty }\int_{0}^{2 N_L}e^{i2\pi
m p}F^{(L)}_{k,k_{0}}(p\, ,t) ~ dp   \\
&=&-\frac{1}{4\pi }\sum_{m=-\infty }^{\infty }\int_{0}^{2\pi}e^{2 i N_L m x- i  t\cos x}\left[ e^{i\left( k+k_{0}\right) x}+e^{-i\left(k+k_{0}\right) x}\right]dx  \\
&\,& + \frac{1}{4\pi }\sum_{m=-\infty }^{\infty }\int_{0}^{2\pi}e^{2 i N_L m x- i  t\cos x}\left[ e^{i\left( k-k_{0}\right) x}+e^{-i\left( k-k_{0}\right) x}\right] dx   \\
&=&-\frac{1}{2}\sum_{m=-\infty }^{\infty } \left( -i\right)
^{k+k_{0}+2 N_L m}J_{k+k_{0}+2 N _L m}\left(t\right)   \\
&\,& -\frac{1}{2}\sum_{m=-\infty }^{\infty }\left( -i\right) ^{-\left( k+k_{0}+2 N_L m\right)
}J_{-\left( k+k_{0}+2 N_L m\right) }\left(t\right)    \\
&\,& + \frac{1}{2}\sum_{m=-\infty }^{\infty }\left( -i\right) ^{k-k_{0}+2 N_L m}J_{k-k_{0}+2 N_L m} \left(t\right)   \\
&\,& +\frac{1}{2}\sum_{m=-\infty }^{\infty }\left( -i\right)
^{-\left( k-k_{0}+2 N_L m\right) }J_{-\left( k-k_{0}+2 N_L m\right) }\left(t\right)  .  \\
\end{eqnarray*}
Using $J_{-n}(x)=\left( -1\right) ^{n}J_{n}\left( x\right) $ we can simplify
the result,
\begin{eqnarray}
U^{(L)}_{k,k_{0}}(t)&=&\left( -i\right) ^{k-k_{0}}\sum_{m=-\infty }^{\infty }\left(
-1\right) ^{N_L m} J_{k-k_{0}+2 N_L m}\left(t\right)  \nonumber \\
&\,& - \left( -i\right) ^{k-k_{0}}\sum_{m=-\infty }^{\infty }\left( -1\right)^{k_{0}+ N_L m}J_{k+k_{0}+2 N_L m}\left( t\right).
\end{eqnarray}
For the case $k_{0}=1$ (i.e.~the situation of a single initially flipped
spin at the left edge that we are interested in), we get
further simplification due to the identity $J_{n-1}(t)+J_{n+1}(t)=2\left( 
n/t\right) J_{n}(t)$,
\begin{eqnarray}
\label{eq:Linear-chain-Propagator-in-terms-of-Bessel-functions}
U^{(L)}_{k,1}(t)=\frac{2\left( -i\right) ^{k-1}}{t}\sum_{m=-\infty }^{\infty
}\left( -1\right) ^{N_L m}\left( k+2 N_L m\right) J_{k+ 2 N_L m}(t ),
\end{eqnarray}
for all $1\leq k \leq
N_L-1$. Eq.(\ref{eq:Linear-chain-Propagator-in-terms-of-Bessel-functions})
is an exact expression that satisfies the initial condition
$U^{(L)}_{k,1}(0)=\delta_{k,1}$. At times less or equal to $N_L$ (a single
propagation from left to right), $U_{k,1}^{(L)}$ can be well approximated by
the term $m=0$,
\begin{eqnarray}
U^{(L)}_{k,1}(t) \simeq \frac{2\left( -i\right) ^{k-1}}{t}k J_{k}(t) ,
\end{eqnarray}
owing to the fact that $ \left|J_k\left(t\right)\right| \ll 1$ for
$\left|t\right| \ll k $ and $k\gg 1$. 
Fig.\ref{fig:Linear-chain-Comparison-Exact-Bessel-Approximated-Bessel-One-term}
shows an example of the time dependence of $U^{(L)}_{k,1}$ for
$N_L=20,k=5$. Visible disagreement of the numerically exact propagator and
the single-Bessel function approximation develops only for $t \gtrsim 27$. 
Each additional term in
(\ref{eq:Linear-chain-Propagator-in-terms-of-Bessel-functions}) increases
the range of $t$ where the Bessel approximation is valid by $N_L$. E.g.,
with $N_L=20,k=5$, adding two more Bessel functions  (terms $m=\pm 1$)  we
get the continuation of the plot in
Fig.\ref{fig:Linear-chain-Comparison-Exact-Bessel-Approximated-Bessel-One-term},
with a visible deviation of the approximation from the exact result only at
$t\gtrsim 65$ (see
Fig.
\ref{fig:linear-chain-Comparison-Exact-Bessel-Approximated-Bessel-Three-terms}).
Physically, these terms correspond to waves reflected from the right and
left edges, which explains their irrelevance for times when the spin waves
have not yet reached the corresponding boundaries. 
\subsection{Semiclassical propagator: WKB approximation}
\begin{figure}[ht]
\includegraphics[width=8cm]{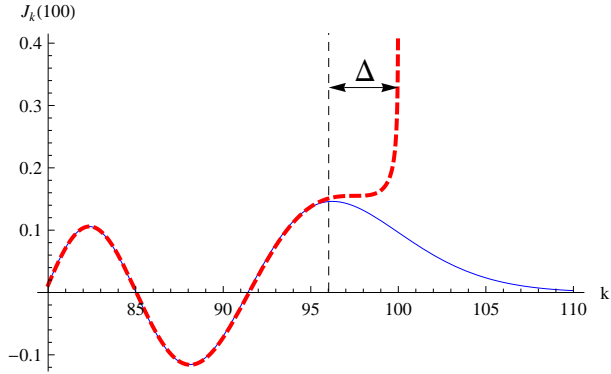} 
\caption{(Color online) Comparison of exact Bessel function (continuous blue
  line) and its ``WKB
  approximation'' (dashed red line), Eq.(\ref{JWKB})}
\label{fig:Break-of-WKB}
\end{figure}

In section \ref{sec:Linear_Mean_Polarization} we will attempt to
obtain a closed analytical formula for the time dependent polarization. In
order to do so, another simplification of $U_{k,1}^{(L)}$ is in order. In
fact, the leading term $m=0$ in
(\ref{eq:Linear-chain-Propagator-in-terms-of-Bessel-functions}), that gives
$U^{(L)}_{k,1}(t) = \frac{2\left( -i\right) ^{k-1}}{t}k J_{k}(t)$, valid for
$t \lesssim N_L$, lends itself to further approximation. From
\cite{Abramowitz_Stegun_Handbook} (leading term of 9.3.15), we can obtain a
``WKB approximation''  for the Bessel function, 
\begin{eqnarray}
\label{eq:Bessel-WKB}
J_k^{WKB}(t) &=&\sqrt{\frac{2}{\pi}}\frac{1}{\sqrt[4]{t^2-k^2}}\cos\left(\phi_k(t)\right) \mbox{ for } k<t , \label{JWKB}\\
\phi_k(t)&=& k\arccos \left(\frac{k}{t}\right)-\sqrt{t^2-k^2}+\frac{\pi}{4} ~ .
\end{eqnarray}
This approximation is commonly called the Debye approximation \cite{WeissteinDebye}. The name
``WKB approximation'' is motivated by the 
fact that the Bessel function is approximated by a sum of two exponential
functions with slowly varying amplitude and phase, in close analogy to the
well-known WKB approximation method. The corresponding approximation of
$U_{k,1}^{(L)}(t)$ can also be obtained from a semiclassical solution of the
Schr\"odinger equation, i.e.~a Van Vleck propagator. This will be presented
in \ref{seq:Circular-chain-Van-Vleck-Approach} for circular chains. The WKB
approximation breaks down near the classical turning point,  which
corresponds here to $k=t$ (see Fig.\ref{fig:Break-of-WKB}). For $k>t$,
$\phi_k(t)$ becomes complex and Eq.(\ref{JWKB}) has to be replaced by an
exponentially decaying function. In the vicinity of the turning point, a
uniform approximation is called for which interpolates smoothly between the
two regimes. 
Bessel functions can be approximated near the turning
point by an Airy function (\cite{Abramowitz_Stegun_Handbook}: 9.3.23), 
\begin{eqnarray}
J_k(t) \approx \left(\frac{2}{t}\right)^{\frac{1}{3}}Ai(-z), ~ z = \left(\frac{2}{t}\right)^{\frac{1}{3}} (t-k) ~.
\end{eqnarray}
This allows to determine precisely the domain in which the WKB approximation is valid, and where it is not.
Fig. \ref{fig:Break-of-WKB} shows the comparison for $t = 100$ between exact
Bessel function and its WKB approximation. The two plots practically
coincide up to the last maximum of $J_k(t)$ as function of  $k$, observed at
$k_m = t- \Delta$. After that the WKB approximation tends to infinity like
$(t-k)^{-\frac{1}{4}}$ whereas the exact $J_k(t)$ quietly goes to zero. The
maximal allowed approach of $k$ to $t$ in the WKB estimate is given by the
position of the last maximum of the Airy function $Ai(-z)$ which takes place
at $z_m=1.01879$  and is equal to $Ai(-z_m)=0.535$. The width of the ``bad''
region where WKB is senseless is thus given by $\Delta =
t-k_m=\left(\frac{t}{2}\right)^{\frac{1}{3}}z_m \approx 0.8
t^{\frac{1}{3}}$. It grows with time, but slowly compared to $t$. 
\subsection{Mean polarization}
\label{sec:Linear_Mean_Polarization}
The total polarization of the chain is defined as $ \displaystyle P^{(L)}  =  \sum_{k=1}^{N_L-1} Z_k $.
In a basis state $\left| \Lambda_k \right \rangle$, we have the mean
polarization $\left\langle \Lambda_k\right|P^{(L)} \left|
\Lambda_k \right\rangle = N_L-2k $. 
If we put the system initially in the state $\left| \Lambda_1 \right
\rangle$, we obtain the time dependent mean polarization  
\begin{eqnarray}
\label{eq:Linear-chain-Exact-polarization}
\langle P^{(L)}\left(t\right)\rangle &=&\sum_{k=1}^{N_L-1}\left(N_L-2k\right)\left|U_{k,1}^{(L)}(t)\right|^2 ~.
\end{eqnarray}
Now let us substitute the WKB approximation of the propagator into the
expression of the mean polarization. We are allowed to do so only in the
``good'' interval where the summands are smooth functions of $k$ and the
sums can be approximated by integrals. Neglecting the exponentially small
terms for $k\ge t$, we are thus led to
\begin{eqnarray}
\label{eq:Linear-chain-WKB-Polarization}
\left \langle P_{WKB}^{(L)}(t) \right \rangle  &=& \sum_{k<t} (N_L-2k) \left|U^{(L)}_{k,1}(t) \right|^2 \nonumber \\
 &=& \sum_{k=1}^{[t-\Delta]} (N_L-2k) \left|U^{(L)}_{k,1}(t) \right|^2 \nonumber \\
&& + \sum_{k=[t-\Delta]}^{[t]} (N_L-2k) \left|U^{(L)}_{k,1}(t) \right|^2 \nonumber \\
&=& \frac{8}{\pi t^2}\sum_{k=1}^{[t-\Delta]} \frac{k^2  (N_L-2k)}{\sqrt{t^2-k^2}} \frac{1+\cos[2 \phi_k(t)]}{2} \nonumber \\
&& + \sum_{k=[t-\Delta]}^{[t]} (N_L-2k) \left|U_{k,1}(t) \right|^2  \nonumber \\
&\simeq& \frac{4}{\pi t^2}   \int_{0}^{t}  (N_L-2k)\frac{k^2}{\sqrt{t^2-k^2}} dk \nonumber\\
&& - \frac{4}{\pi t^2}   \int_{t-\Delta}^t  (N_L-2k)\frac{k^2}{\sqrt{t^2-k^2}} dk \nonumber \\
&& + \frac{4}{\pi t^2}\sum_{k=1}^{[t-\Delta]} \frac{k^2 (N_L-2k)}{\sqrt{t^2-k^2}} \cos(2\phi_k(t)) \nonumber \\
&& + \sum_{k=[t-\Delta]}^{[t]}(N_L-2k)\left|U^{(L)}_{k,1}(t) \right|^2 ~,
\end{eqnarray} where $[x]$ denotes the integer part of $x$.
The first integral in (\ref{eq:Linear-chain-WKB-Polarization}) is proportional to $t$, 
\begin{eqnarray}
\label{eq:Linear-chain-WKB-Polarization-Linear-term}
\left \langle P_{WKB}^{(L)}(t)\right \rangle \simeq N_L -\frac{16 t }{3\pi} ~.
\end{eqnarray}
Let us make an estimate of the remaining terms. The second integral $
\frac{1}{t^2}\int_{t-\Delta}^{t}\ldots$ is of the order $ \sqrt{t \Delta}\approx
t^{2/3}$ which is small compared to $t$ when $t \rightarrow \infty$. The sum
containing $\cos\left[2 \phi_k(t)\right]$ contains many summands of both
signs which mutually almost cancel and can be neglected (rigorous estimates can be made using
the Poisson summation formula). Finally, the last sum can be estimated as the
number of summands $\Delta \propto t^{1/3}$, times the maximal value of the
summand $(N_L-2t) \left[ 2 \left(\frac{2}{t}\right)^{1/3} Ai(-z_m)\right]^2
\propto t^{1/3}$. The result is again $ \propto t^{2/3}$ and can again  be
neglected compared to $t$ when $t$ is very large. 
\begin{figure}[ht]
\includegraphics[width=8cm]{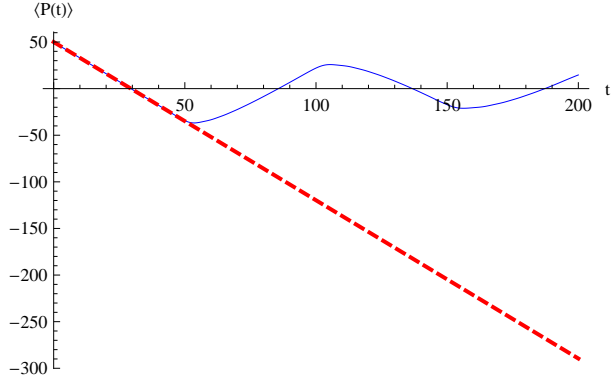}
\caption{(Color online) Comparison of mean polarizations: exact result (Eq.(\ref{eq:Linear-chain-Exact-polarization}), continuous blue line) versus linear term of WKB approximation (Eq.(\ref{eq:Linear-chain-WKB-Polarization-Linear-term}), dashed red line), for $N_L=50$}
\label{fig:Linear-chain-Comparison-Exact-Polarization-WKB-Linear-term}
\end{figure}
Thus the estimates for the relative errors decay like $t^{-1/3}$  as $t
\rightarrow \infty$, and
Eq.(\ref{eq:Linear-chain-WKB-Polarization-Linear-term}) is therefore the
leading behavior of the WKB approximation. While this approximation is
rather crude, it 
nevertheless  shows very good agreement with the exact result for times
small enough not to allow spin waves to reach the end of the chain. E.g.,
for $N_L = 50$ and $t=40$ the Bessel representation (which is basically exact
here), gives $ P(t)=-17.922$ whereas the WKB result is  $N_L-\frac{16
}{3\pi}t=-17.9601$, i.e.~reproduces the Bessel result with 3 digits. In
Fig.\ref{fig:Linear-chain-Comparison-Exact-Polarization-WKB-Linear-term} we
compare the exact mean polarization
(\ref{eq:Linear-chain-Exact-polarization}) with its semiclassical
approximation (\ref{eq:Linear-chain-WKB-Polarization-Linear-term}) for a
chain of $50$ spins. 
From Eq.(\ref{eq:Linear-chain-WKB-Polarization-Linear-term}) we read off a
constant rate of spin flips (corresponding here to the speed of propagation of a single spin-wave front) for times $t \apprle N_L$, $ \left| v^{(L)}
\right|  = \frac{16}{3 \pi }$. 
\section{Circular chain}\label{sec.cc}
There are several ways of closing the linear chain (\ref{eq:Linear-chain-Hamiltonian}) of $N_L$ spins to a ring:
\begin{enumerate}
\item By identification of $S_{N_L + 1}$ and $S_1$: in this system $S_2,
  \ldots, S_{N_L}$ are subject to spin flipping (see
  Eq.(\ref{eq:Circular-chain-Hamiltonian-Closing-by-Spin-Identification})
  below); 
\item By identification of $S_{N_L + 1}$ and $S_1$ and introduction of an
  additional interaction term  containing $X_1$ in (\ref{eq:Linear-chain-Hamiltonian})
  that allows $S_1$ to flip: in this system $S_1, \ldots, 
  S_{N_L}$ are subject to spin flipping (see Eq.(\ref{eq:Circular-hamiltonian-Closing-by-adding-a-coupling}) below); 
\item By introducing a different coupling between the first spins and the
  last spins of the linear chain: in section
  \ref{seq:Closing-by-particular-coupling}, we will introduce a
  particular four-spin coupling between $S_{N_L-1}, S_{N_L}, S_1, S_2$ .
\end{enumerate}
We will now investigate the dynamics in these different chains. $S_1$ will designate always the spin chosen as the reference spin from which all the others are numbered (from $S_2$ to $S_{N_L}$ turning clockwise on the chains).
\subsection{Closing by identification of spins $S_{N_L+1}$ and $S_1$}
\label{seq:Circular-Chain-Closing-by-Spin-identification}
The simplest way of closing a linear chain consists in identifying the first
and the last spin. When we talk about circular chains we will call the
number of spins $N_C$. For the present subsection we consider $N_C=N_L$, such that $S_{N_C+1} =S_1$. This imposes a corresponding boundary condition on the
wave function, but also implies an extra term in the hamiltonian since identifying $S_1$ with
$S_{N_C+1}$ allows to flip spin $S_{N_C}$ depending on the state of
$S_{N_C-1}$ and $S_{N_C+1}=S_1$. Thus we have to add the term $k=N_C$ to
$H_1^{(L)}$, i.e. 
\begin{eqnarray}
\label{eq:Circular-chain-Hamiltonian-Closing-by-Spin-Identification}
H_1^{(C)} = \frac{1}{4}\sum_{k=2}^{N_C}X_k \left( 1_{k-1} \otimes 1_{k+1} - Z_{k-1}\otimes Z_{k+1}\right).\label{H1c}
\end{eqnarray}
Note that $S_1$ still never flips as there is no term containing $X_1$. In
section \ref{seq:Circular-chain-Closing-by-adding-a-coupling} we will
consider the situation of complete rotational symmetry implemented already
on the level of the hamiltonian, where the introduction of yet another flip
term also allows $S_1$ to flip. 
The system described by
Eq.(\ref{eq:Circular-chain-Hamiltonian-Closing-by-Spin-Identification})
evolves inside a subspace whose $N_C(N_C-1)/2$ basis states $\left|
\chi_{I,J}\right \rangle $ ($I \in [1, N_C-1]$ and $J \leq I$) can be
arranged in the form of a triangle, 
\begin{eqnarray}
\label{eq:Triangle-of-States}
&& \left|\chi_{1,1}\right\rangle \nonumber \\ 
&&  \left|\chi_{2,1}\right\rangle , \left|\chi_{2,2}\right\rangle \nonumber  \\
&&  \left|\chi_{3,1}\right\rangle  ,  \left|\chi_{3,2}\right\rangle  , \left|\chi_{3,3}\right\rangle  \nonumber \\
&& \vdots  \hspace{90pt}\ddots \nonumber \\
&& \left| \chi_{N_C-1,1}\right\rangle,~ \ldots\ldots\ldots ~,\left|\chi_{N_C-1,N_C-1}\right\rangle ~.
\end{eqnarray}
We define $\left| \chi_{I,J} \right\rangle $ as the state containing $I$
consecutive spins down (all the others are up), with $J-1$ spins down at the
left of $S_1$, $S_1$ down, and $I-J$ spins down at the right of $S_1$
(imagine $S_1$ at the top of the chain). For instance,
$\left|\chi_{1,1}\right\rangle = \left| \downarrow_1 \uparrow_2 \ldots
\uparrow_{N_C} \right\rangle $, $\left| \chi_{2,1}\right\rangle = \left|
\downarrow_1 \downarrow_2 \uparrow_3 \ldots \uparrow_{N_C} \right\rangle $, 
$\left|\chi_{2,2}\right\rangle = \left| \downarrow_1 \uparrow_2 \ldots
\uparrow_{N_C-1} \downarrow_{N_C} \right\rangle $, 
$\left|\chi_{3,1}\right\rangle = \left| \downarrow_1 \downarrow_2
\downarrow_3 \uparrow_4 \ldots \uparrow_{N_C} \right\rangle $,
$\left|\chi_{3,2}\right\rangle = \left| \downarrow_1  \downarrow_2
\uparrow_3 \ldots \uparrow_{N_C-1} \downarrow_{N_C} \right\rangle $,
$\left|\chi_{3,3}\right\rangle = \left| \downarrow_1  \uparrow_2 \ldots
\uparrow_{N_C-2} \downarrow_{N_C-1} \downarrow_{N_C} \right\rangle $, etc. 
 
$H_1^{(C)}$ couples a state $\left|\chi_{I,J}\right\rangle$ to at most four
neighbors in the triangle: the one above, the one above to the left, the one
below, and the one below to the right. Depending on the region in the
triangle, not all four of the states exist. A state is coupled to exactly
those of the four states that do exist. 
In order to obtain a matrix representation of $H_1^{(C)}$, we order the
states $\left|\chi_{I,J}\right\rangle$ in the order $
\left|\chi_{1,1}\right\rangle ,  \left|\chi_{2,1}\right\rangle ,
\left|\chi_{2,2}\right\rangle ,  \left|\chi_{3,1}\right\rangle \ldots$,
i.e.~we introduce a single label $L(I,J)=I(I-1)/2 + J$. Therefore, top state
$1$ couples to states $\left\{ 2,3\right\}$ , lower left hand corner state
$L=\frac{(N_C-1)(N_C-2)}{2}+1$ couples to state $  L-N_C+2$, lower right hand
corner state $L=\frac{(N_C-1)(N_C-2)}{2} + N_C-1$ couples to state $L-N_C+1$,  left
border states $L=\frac{I(I-1)}{2}+1$ with $I\in\left[2, N_C-2\right]$  couple
to states $\left\{L-I+1,L+I,L+I+1 \right\}$,   
right border states $L=\frac{I(I-1)}{2}+I $ with $I\in\left[2, N_C-2\right]$
couple to states $\left\{L-I ,L+I,L+I+1\right\}$,    
base states $L=\frac{(N_C-1)(N_C-2)}{2}+ J$ with $J=\left[ 2,N_C-2\right]$
couple to states $\left\{L-N_C+1,  L-N_C+2 \right\}$,  and inside states
$L=\frac{I(I-1)}{2}+ J$ with $I\in\left[3, N_C-2\right]$ and $ 1 <J <I$ couple
to states $  \left\{L-I , L-I+1, L+I,L+I+1\right\}$. 
For example, the matrix representation of $H_1^{(C)}$ for $N_C = 5$ ($\frac{N_C(N_C-1)}{2} = 10$ basis states) reads
\begin{eqnarray}
H_1^{(C)} =  \frac{1}{2}\left(
\begin{array}{cccccccccc}
   0&1&1&0&0&0&0&0&0&0 \\
   1&0&0&1&1&0&0&0&0&0 \\
	1&0&0&0&1&1&0&0&0&0 \\
   0&1&0&0&0&0&1&1&0&0 \\
   0&1&1&0&0&0&0&1&1&0 \\
   0&0&1&0&0&0&0&0&1&1 \\
   0&0&0&1&0&0&0&0&0&0 \\
   0&0&0&1&1&0&0&0&0&0 \\
   0&0&0&0&1&1&0&0&0&0 \\
   0&0&0&0&0&1&0&0&0&0 \\
\end{array}
\right) ~. 
\end{eqnarray}
The general structure of the matrix corresponding to a system of $N_C$
spins can be easily derived. The matrix is real and symmetric, and we
need to consider only the upper right triangle  $(H_1^{(C)})_{L,L'}$  with
$L'>L$. To fill this part of the matrix we have to know for a state of the
basis to which state below in the triangle (\ref{eq:Triangle-of-States}) it
is connected. Each state from the first line
to the last but one of the triangle is only connected to two consecutive
states below it in the triangle (the state directly below and the one below
to the right). Therefore, in each line of the upper right part of
$H_1^{(C)}$ there are only two consecutive non zero elements. Given two
states $\chi_L$ and $\chi_{L+1}$ of a same line of the triangle, there is
one common state among the states coupled to $\chi_L$ and the ones coupled
to $\chi_{L+1}$. That explains that we observe staircase-structures
extending over $1, 2, 3, \ldots$ rows . Since there are $N_C-1$ lines in
(\ref{eq:Triangle-of-States}) and each line with the exception of the last
one gives rise to a staircase, there are altogether $N_C-2$ staircases. From
one staircase structure to the next, there is a shift of one line and one
column, since the first state of a line is not connected to the last state of
the preceding one.  

A brief remark is in order about the global structure of the distribution of
non-zero elements over the matrix. Let us consider for each
staircase-structure the first element, i.e.~$(H_1^{(C)})_{1,2}$,
$(H_1^{(C)})_{2,4}$, $(H_1^{(C)})_{4,7}$, $\ldots$~ The coordinates (line,
column) of these elements are given by the index $L$ for the elements of the
left border in (\ref{eq:Triangle-of-States}), $X_I = L =
\frac{I(I-1)}{2}+1$, and the element below it, $Y_I =
X_I+I=\frac{I(I+1)}{2}+1$. If we plot the $(X_I,Y_I)$ points in a plan, we
obtain a line whose slope $\frac{Y_{I+1}-Y_I}{X_{I+1}-X_I} = 1+\frac{1}{I}$
converges for $I \rightarrow \infty$ to $1$. This means that for $N_C \gg
1$, the matrix representation of $H_1^{(C)}$ is very close to a matrix
having four non zero diagonals parallel to the main diagonal, but a final
curvature of the ``off-diagonals'' remains for all finite $I$. In the
next section we see that straight lines of off-diagonal matrix elements over
the entire matrix, parallel to the main diagonal, are obtained  with the second
way of closing the chain.  
Lacking a viable analytical technique for diagonalizing $H_1^{(C)}$, we
diagonalize the hamiltonian numerically, and derive the propagator. The
results will be compared to those of the linear chain in section
\ref{seq:Comparison-between-Linear-chain-Circular-Chains}. 
\subsection{Completely periodic chain}
\label{seq:Circular-chain-Closing-by-adding-a-coupling}
Another way of closing the system is to impose periodic boundary conditions
$S_{N_C+k} = S_k$, $k \in \left[1,N_C\right]$. This amounts to subjecting
also $S_1$ to flip by adding an additional flip term $X_1$ controlled by
$S_{N_C}$ and $S_2$. The hamiltonian of this system thus reads 
\begin{eqnarray}
\label{eq:Circular-hamiltonian-Closing-by-adding-a-coupling}
H_2^{(C)} &=& \frac{1}{4}\sum_{k=1}^{N_C} X_k \left( 1_{k-1} \otimes 1_{k+1} - Z_{k-1}\otimes Z_{k+1}\right) ~ .
\end{eqnarray}
The physical important point is that in this system $S_1$ is flipped if its
nearest-neighbors are of opposite signs ($S_{N_C}$ and $S_2$), whereas in
section \ref{seq:Circular-Chain-Closing-by-Spin-identification}, $S_1$ is
never subjected to spin flipping and always remains in its initial down
state. As a consequence, the number of basis states that are dynamically
accessible is doubled. Indeed, allowing $S_1$ to be flipped, gives access to
basis states related to the $\left| \chi_L\right \rangle$ by flipping all
spins. They form an ensemble of states having $k$ consecutive spins down
($k \in \left[1,N_C-1\right]$), where in addition, for each  $k \in
\left[1,N_C-1\right]$, we have the freedom of choosing the starting point of
the sequence of flipped spins among the $N_C$ spins. So the dimension of the
accessible Hilbert space is now $N_C( N_C-1)$. 
\begin{figure}[ht]
\includegraphics[width=9cm]{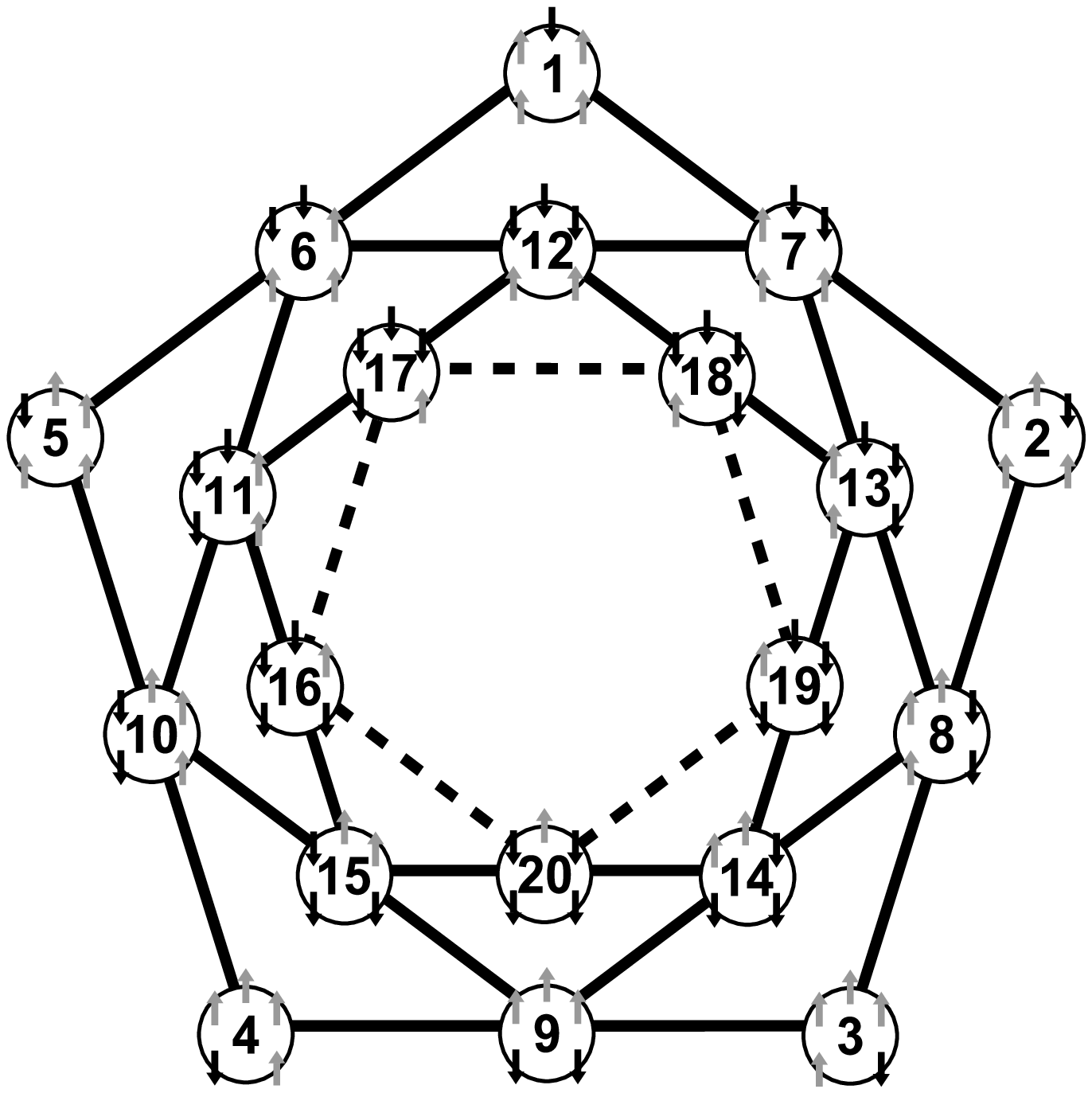} 
\caption{(Color online) Coupling of states introduced by $H_2^{(C)}$, Eq.(\ref{eq:Circular-hamiltonian-Closing-by-adding-a-coupling}), for $N_C=5$}
\label{fig:states_connections_for_H2}
\end{figure}
It is interesting to note that the couplings between the basis states can be obtained by arranging the states on the  vertices of $(N_C-1)$
nested $N_C$-sided polygons as can be observed in
Fig.\ref{fig:states_connections_for_H2}  for $N_C=5$. A full line
represents the coupling between two states. We label the states as can be
seen in  Fig.\ref{fig:states_connections_for_H2}: we follow a polygon
until all of its states have been accounted for. Then we move to the next
polygon further inside by moving to its first vertex on the center of the last link of the previous polygon, and so on. Using this
numbering, we obtain a matrix representation of $H_2^{(C)}$, with a simple
structure. For example for $N_C=4$ we obtain 
\begin{eqnarray}
\label{eq:hamiltonian-H2-N4}
H_2^{(C)} =  \frac{1}{2}\left(
\begin{array}{cccccccccccc}
   0&0&0&0&1&1&0&0&0&0&0&0 \\
   0&0&0&0&0&1&1&0&0&0&0&0 \\
   0&0&0&0&0&0&1&1&0&0&0&0 \\
   0&0&0&0&1&0&0&1&0&0&0&0 \\
   1&0&0&1&0&0&0&0&1&1&0&0 \\
   1&1&0&0&0&0&0&0&0&1&1&0 \\
   0&1&1&0&0&0&0&0&0&0&1&1 \\
   0&0&1&1&0&0&0&0&1&0&0&1 \\
   0&0&0&0&1&0&0&1&0&0&0&0 \\
   0&0&0&0&1&1&0&0&0&0&0&0 \\
   0&0&0&0&0&1&1&0&0&0&0&0 \\
   0&0&0&0&0&0&1&1&0&0&0&0 \\
\end{array}
\right) ~. 
\end{eqnarray}
The general structure corresponding to a system of $N_C$ spins can again be
easily derived. Since the matrix is real and symmetric, we need to consider
only the upper right triangle  $(H_2^{(C)})_{L,L'}$  with $L'>L$. We observe
two neighboring off-diagonals, parallel to the main diagonal, and a few non zero elements next to the
(vanishing) main diagonal. The non zero terms directly next to the main
diagonal are the elements  $ (H_2^{(C)})_{k N_C,k N_C + 1}$ with
$k\in\left[1, N_C-2\right]$. The first diagonal extends from
  $(H_2^{(C)})_{1,N_C+1}$ to $(H_2^{(C)})_{N_C(N_C-2),N_C(N_C-1)}$ (all
  elements equal to one), the second diagonal goes from
  $(H_2^{(C)})_{1,N_C+2}$ to $(H_2^{(C)})_{N_C(N_C-2)-1,N_C(N_C-1)}$ (all
  elements equal to one except the $(k N_C)^{th}$ with $k\in [1,N_C-3]$). 
The fact that the two most straightforward ways of closing the chain lead to
a relevant Hilbert space of dimension of ${\cal O}(N_C^2)$ instead of ${\cal O}(N_L)$, is
rather interesting and  will be explored further below. At the same time it
would be interesting to find out whether there exist other ways of closing
the chain that lead to dynamics resembling as closely as possible, the
dynamics of the linear chain, and in particular to a relevant Hilbert space
of dimension of ${\cal O}(N_C)$. In the following two
  subsections we present such a closure and
profit from the machinery developed in section \ref{seq:Linear-chain} to
study the corresponding dynamics. 
\subsection{Closing by a particular coupling between $S_{N_C-1}$, $S_{N_C}$, $S_1$, and $S_2$}
\label{seq:Closing-by-particular-coupling}
We now study a particular way of closing the linear chain for which, as we
are going to see, the system evolves in a Hilbert space  whose dimension
scales only linearly with the size of the chain. We consider a system that
has the same nearest-neighbors interactions as $H_2^{(C)}$  for spin $S_2$
to spin $S_{N_C-1}$, and an additional particular coupling for spins
$S_{N_C-1}$, $S_{N_C}$, $S_1$ and $S_2$ that leads to the following action:
if spin  $S_{N_C-1}$ and  spin $S_2$ are the same, then spin $S_{N_C}$ and
spin $S_1$ are flipped, otherwise nothing is done. The hamiltonian of such a
system is 
\begin{eqnarray}
\label{eq:Circular-chain-Closed-by-particular-coupling-Hamiltonian}
H_3^{(C)} &=& \frac{1}{4}\sum_{k=2}^{N_C-1}X_k \left( 1_{k-1} \otimes 1_{k+1} - Z_{k-1}\otimes Z_{k+1}\right) \nonumber \\
&&+ \frac{1}{4} X_{N_C}X_1 \left(1_{N_C-1} 1_2 + Z_{N_C-1}Z_2\right)~ .
\end{eqnarray}
Note the small difference compared to $H_2^{(C)}$. As before, we
denote $\left| \chi_k \right \rangle  = \left|\downarrow_1 \ldots
\downarrow_k\uparrow_{k+1} \ldots \uparrow_{N_C} \right\rangle  =\left|
\Lambda_k\right\rangle$ the state 
where spins $S_1$ to $S_k$ are down and all the others up for $ k \in\left[1
  \, , N_C-1\right]$. Then we define $ \left| \chi_{N_C+k-1} \right\rangle =
\left|\uparrow_1\ldots\uparrow_k \downarrow_{k+1} \ldots \downarrow_{N_C}
\right\rangle  = \left| \overline{\Lambda_k} \right \rangle $, $ k
\in\left[1 \, , N_C-1\right] $, the state where spins $S_1$ to $S_k$ are up,
all the others down. The $\left|\overline{\Lambda_k}\right\rangle $ are the
mirror states of  $\left| \Lambda_k\right\rangle$ in
the sense that all 
spins are flipped. 
Let us suppose again that the system starts in the state $\left| \chi_1
\right \rangle = \left|\downarrow_1 \uparrow_2 \ldots \uparrow_{N_C}
\right\rangle $. This state couples to  $\left| \chi_2 \right \rangle   =
\left|\downarrow_1 \downarrow_2\uparrow_3 \ldots \uparrow_{N_C}
\right\rangle $ and $\left|\chi_{2(N_C-1)} \right \rangle = \left|\uparrow_1
\uparrow_2\ldots\uparrow_{N_C-1} \downarrow_{N_C} \right\rangle$. For
general $ k \in \left[2\, , N_C - 2\right]$, $\left| \chi_k\right \rangle $
couples to  $\left| \chi_{k-1}\right \rangle $ and $\left|\chi_{k+1} \right
\rangle $ and $\left| \chi_{N_C+k-1}\right \rangle
=\left|\overline{\Lambda_k}\right\rangle $ couples to
$\left|\overline{\Lambda_{k-1}}\right\rangle$ and
$\left|\overline{\Lambda_{k+1}}\right\rangle$. Thus, at the end of the
chain, the system branches over into the mirror states, which in turn lead
back to the original states and therefore give rise to a closed basis of
$2(N_C-1)$ states. We note, however, that neither the state with all spins
up nor the state with all spins down is any longer an eigenstate of
$H_3^{(C)}$, with consequences for the spin amplification discussed in
Sec.\ref{seq:Comparison-between-Linear-chain-Circular-Chains}. The matrix of
the hamiltonian in the basis of the $\left| \chi_k \right \rangle$, for $k
\in \left[1,2(N_C-1)\right]$ is  
\begin{eqnarray}
\label{eq:Circular-chain-Closing-by-particular-coupling-Matrix-representation-of-hamiltonian}
H_3^{(C)}=  \frac{1}{2}\left(
\begin{array}{ccccccc}
   0&1&0&\ldots&\ldots&0&1\\
   1&0&1&\ddots&~&~&0\\
   0&1&0&\ddots&\ddots&~&\vdots\\
\vdots&\ddots&\ddots&\ddots&\ddots&\ddots&\vdots\\
\vdots&~&\ddots&\ddots&0&1&0\\
0&~&~&\ddots&1&0&1\\
1&0&\ldots&\ldots&0&1&0\\
\end{array}
\right) ~.
\end{eqnarray}
This is a well-known hamiltonian of a regular chain closed into a ring. Its
eigenvalues and eigenvectors are given by 
\begin{eqnarray}
\label{eq:HC3_Eigensystem}
&&\lambda_p^{(C)} = \cos[\frac{p \, \pi}{N_C-1}] \mbox{ and }\nonumber \\
&&\left|\Phi_p^{(C)}\right\rangle =\frac{1}{ \sqrt{2(N_C-1)}}\sum_{k=1}^{2(N_C-1)}e^{i\frac{p \pi k }{N_C-1}}\left|\chi_k\right\rangle 
\end{eqnarray}
for $p \in \left[1,2(N_C-1)\right]$. Finally, we obtain the expression of the propagator
\begin{eqnarray}
U_{k,k_0}^{(C)}\left(t \right) &=&\left\langle \chi_k \right\vert
e^{-i H_3^{(C)} t}\left\vert \chi_{k_{0}}\right\rangle \nonumber \\
&=&\sum_{p=1}^{2(N_C-1)}\left\langle \chi_k\right\vert \left. \Phi^{(C)}
_{p}\right\rangle e^{-it\lambda _{p}^{(C)}}\left\langle \Phi^{(C)}_{p}\right\vert
\left. \chi_{k_{0}}\right\rangle  \nonumber \\
&=&\sum_{p=1}^{2(N_C-1)}F^{(C)}_{k,k_{0}}(p \, , t)\\
\nonumber
\end{eqnarray}
with $F^{(C)}_{k,k_{0}}(p \, , t) \equiv \frac{1}{2(N_C-1)} e^{i \frac{p \pi (k-k_0)}{N_C-1}}e^{-it \cos \frac{p \pi}{N_C-1}}~.$
\subsubsection{Representation of the propagator in terms of Bessel functions}
Using that $ F^{(C)}_{k,k_{0}}(p \ + 2(N_C-1) \, , t)  =F^{(C)}_{k,k_{0}}(p \, , t)$, we can use the Poisson summation formula to find 
\begin{eqnarray}
U_{k,k_0}^{(C)}\left(t \right) &=&\sum_{m=-\infty}^{\infty} \int_0^{2(N_C-1)} e^{2 i \pi m p} F^{(C)}_{k,k_{0}}(p \, , t)\,dp\nonumber \\
&=&\sum_{m=-\infty}^{\infty}(-i)^{k-k_0+2m(N_C-1)}J_{k-k_0+2m(N_C-1)}(t) ~ .
\end{eqnarray}
For the special case where the system starts in the state $\left|\chi_1\right \rangle$, we obtain
\begin{eqnarray}
\label{eq:Circular-chain-H3-Propagator-in-terms-of-Bessel-functions}
U_{k,1}^{(C)}\left(t\right) &=& (-i)^{k-1} \sum_{m=-\infty}^{\infty}(-1)^{m (N_C-1)}J_{k-1+2 m (N_C-1)}\left(t\right) .
\end{eqnarray}
\subsubsection{Semiclassical propagator for  small times ( $t< N_c-1$) }
The Bessel function vanishes exponentially fast if its index is large and
exceeds its argument. Therefore the infinite sum
(\ref{eq:Circular-chain-H3-Propagator-in-terms-of-Bessel-functions}) reduces
in fact to only a few summands significantly different from zero.  In
particular, if $t<N_c-1$ only a single term corresponding to $m=0$ or $m=-1$
survives respectively for $1\le k\le N_c-1$ (states $\Lambda$) and $N_c\le
k\le 2(N_c-1)$ (states $\overline{\Lambda}$),
\begin{eqnarray}
U_{k,1}^{(C)}(t)\approx (-i)^{k-1}J_{k-1}(t)\mbox{, }\,\,\,1\le k\le
N_c-1;\nonumber\\
U_{k,1}^{(C)}(t)\approx (-1)^{N_C-1}(i)^{k-1}J_{2N_c-k-1}(t)\mbox{, }\,\,\,N_c\le k\le 2(N_c-1)\,.\nonumber
\end{eqnarray}
If $1\ll k$ (respectively $1\ll 2N_c-k-1$), these Bessel functions can be
replaced by their WKB approximation (\ref{eq:Bessel-WKB}). 
\subsubsection{Van Vleck approach}
\label{seq:Circular-chain-Van-Vleck-Approach}
Before considering the polarization dynamics, we present an alternative and
more physical approach for calculating the 
semiclassical propagator for $k\in \left[1,N_C-1 \right]$ and $k<t+1$. The
same problem as before can be approached by deriving the classical
hamiltonian associated to the system; the method and its generalization to
chains with slowly changing parameters are described in the review
\cite{PBraun93}. The action of $H_3^{(C)}$ on a basis 
state, $H_3^{(C)} \left| \chi_k \right \rangle = \frac{1}{2} \left( \left|
\chi_{k-1} \right \rangle+\left| \chi_{k+1} \right \rangle\right)$, can be
rewritten in terms of shift operators, if we upgrade the index $k$ to a
continuous variable, i.e.~$H_3^{(C)} \left| \chi(k) \right \rangle =
\frac{1}{2} \left( \left| \chi(k-1) \right \rangle+\left| \chi(k+1) \right
\rangle\right) = \frac{1}{2} \left( e^{-\partial/ \partial_k} \left| \chi(k)
\right \rangle+e^{\partial/ \partial_k}\left| \chi(k) \right \rangle\right)
$. This is a reasonable approach in the limit of very long chains, $N_C
\rightarrow \infty$, which can be considered the classical limit. If we
denote $\phi = -i \partial/ \partial_k$ the momentum canonically conjugate
to the coordinate $k$, we can write  $H_3^{(C)} \left| \chi(k) \right
\rangle = \frac{1}{2} \left( e^{-i \phi} +e^{i \phi} \right) \left|
\chi(k)\right \rangle  = \cos\left(\phi\right)\left| \chi(k)\right
\rangle$. In the classical limit we therefore obtain a corresponding
classical hamiltonian $H_3^{Cl} = \cos\left(\phi\right)$. 
Since $H_3^{Cl}$ does not depend on $k$, the momentum is an integral of
motion connected with the energy by $E = \mbox{cte} = \cos(\phi) $. The
trajectory of the motion follows from the canonical equation $
\dot{k}(t)=\frac{\partial H_3^{Cl}}{\partial \phi}=-\sin(\phi)$, hence $k(t)
= k_0-t\sin(\phi)$. The classical action is given by $ \displaystyle
S(k,k_0,t) = \int_{k_0}^{k} \phi(\tilde{k}) d \tilde{k} - E(k,k_0,t) t = (k-k_0)\phi
- E(k,k_0,t) t$. We find two classical paths, $
\phi_1(k,k_0,t)=-\arcsin(\frac{k-k_0}{t})$ and $
\phi_2(k,k_0,t)=\pi+\arcsin(\frac{k-k_0}{t})$. Their associated classical
energies are $E_1(k,k_0,t) = \cos(\phi_1(k,k_0,t))=\frac{1}{t}\sqrt{t^2 -
  (k-k_0)^2}$ and $E_2(k,k_0,t) = \cos(\phi_2(k,k_0,t))=-\frac{1}{t}\sqrt{t^2 -
  (k-k_0)^2}$. Finally, the associated classical actions are given by
$S_1(k,k_0,t) = (k-k_0)\phi_1(k,k_0,t)-E_1(k,k_0,t) t =
-(k-k_0)\arcsin(\frac{k-k_0}{t})-\sqrt{t^2-(k-k_0)^2}$ and $S_2(k,k_0,t)=
(k-k_0)\phi_2(k,k_0,t)-E_2(k,k_0,t) t= (k-k_0)(\pi +
\arcsin(\frac{k-k_0}{t}))+\sqrt{t^2-(k-k_0)^2}$. 
The semiclassical Van Vleck propagator is obtained by summing over all the classical paths,
\begin{eqnarray}
U^{VV}_{k,k_0}(t) = \sum_{\alpha} \sqrt{\frac{1}{2i\pi}\left|\frac{\partial^2 S_{\alpha}}{\partial k \partial k_0}\right|}\exp(i\left(S_{\alpha}(k,k_0,t) - \nu_{\alpha}\frac{\pi}{2}\right)) ~,
\end{eqnarray}  
where $ \nu_{\alpha}$ is the Morse index for the classical path $\alpha$. For the case of the small times considered we have only two classical paths with $ \nu_1 = -1$ and $\nu_2=0$ and therefore,
\begin{eqnarray}
\label{eq:Circular-chain-Van-Vleck-Propagator}
U^{VV}_{k,k_0}(t) = U^{VV,(1)}_{k,k_0}(t)+U^{VV,(2)}_{k,k_0}(t)
\end{eqnarray}
\begin{eqnarray*}
&&U^{VV,(1)}_{k,k_0}(t)=\sqrt{\frac{i}{2\pi\sqrt{t^2-(k-k_0)^2}}}
\exp(-i X)\\
&&U^{VV,(2)}_{k,k_0}(t)=\sqrt{\frac{1}{2 i \pi\sqrt{t^2-(k-k_0)^2}}}
(-1)^{k-k_0} \exp(i X)
\end{eqnarray*} 
with  $X= (k-k_0)\arcsin(\frac{k-k_0}{t})+\sqrt{t^2-(k-k_0)^2}$. It can be
easily checked that (\ref{eq:Circular-chain-Van-Vleck-Propagator}) with
$k_0=1$ is
exactly the same as $U^{WKB}_{k,1}(t)^{(C)} = (-i)^{k-1} J^{WKB}_{k-1}(t)  $
with $J^{WKB}_{k-1}(t)$ given by (\ref{eq:Bessel-WKB}). 
\subsubsection{Mean polarization}
\label{sec:Circular-chain-Mean-polarization} 
The operator for the total polarization is now  $\displaystyle P^{(C)}
= \sum_{k=1}^{2(N_C-1)} Z_k$. Expressed in the basis of the $\left\{ \left|
\chi_k \right \rangle \right\}$, the mean polarization is $\left\langle \chi_k \right| P^{(C)} \left| \chi_k \right\rangle =
(N_C-2k)$ for  $k \in [1,N_C-1]$, and  $\left\langle\chi_k\right|P^{(C)} \left|  \chi_k \right\rangle=
-(3N_C-2k-2)$ for $k \in [N_C,2(N_C-1)]$. If we put the system initially in
the state $\left| \chi_1 \right \rangle$, we obtain the time dependent mean
polarization, 
\begin{eqnarray}
\label{eq:Circular-chain-Basis-ONC-Exact-Propagator}
\langle P^{(C)}(t)\rangle &=& \sum_{k=1}^{N_C-1}\left(N_C-2k\right)\left|U_{k,1}^{(C)}(t)\right|^2 \nonumber \\
&-& \sum_{k=N_C}^{2(N_C-1)}\left(3N_C-2k-2\right)\left|U_{k,1}^{(C)}(t)\right|^2
\end{eqnarray}
As before, the WKB approximation of the mean polarization is obtained by
replacing the exact propagator in Eq.(\ref{eq:Circular-chain-Basis-ONC-Exact-Propagator}) by its WKB
approximation $U_{k,1}^{WKB}(t)^{(C)}$, 
\begin{eqnarray*}
&&\langle P^{(C)}_{WKB}(t)  \rangle \\
&=& \sum_{k<t+1} (N_C-2k)\left| U^{WKB}_{k,1}(t)^{(C)}\right|^2 \\
&-&  \sum_{k>2(N_C-1) - t+1} (3N_C-2k-2)\left| U^{WKB}_{k,1}(t)^{(C)}\right|^2\\
&=&\sum_{k<t+1} (N_C-2k)\left| J^{WKB}_{k-1}(t)\right|^2 \\
&-&  \sum_{k>2(N_C-1) - t+1} (3N_C-2k-2)\left| J^{WKB}_{2(N_C-1)-k+1}(t)\right|^2\\
&=&\sum_{k<t+1} (N_C-2k)\frac{2}{\pi}\frac{\cos^2\left[\phi[k+1,t]\right]}{\sqrt{t^2-(k-1)^2}} \\
&-&  \sum_{k>2(N_C-1) - t+1} \frac{2}{\pi }(3N_C-2k-2) \frac{\cos^2\left[\phi[2(N_C-1)-k+1,t]\right]}{\sqrt{t^2-\left(2(N_C-1)-k+1\right)^2}} ~. \\
\end{eqnarray*}
Making the same kind of calculation as before and neglecting in the limit
$N_C \gg 1$ the terms in the vicinity of the turning point (see the
discussion about the estimation of remaining terms neglected for the
expression of mean polarization in the case of the linear chain in section
I.D), we finally obtain for the linear contribution of the mean
polarization 
\begin{eqnarray}
\label{eq:Circular-chain-WKB-polarization-Linear-term}
\left \langle P_{WKB}^{(C)}(t) \right \rangle  \simeq (N_C-1) -\frac{4 t }{\pi}.
\end{eqnarray}
We read off a rate of spin flips $\left| v^{(C)}\right| =
\frac{4}{\pi}$. It is reduced by $25\%$ compared to the one obtained for the
linear chain, 
$\left| v^{(L)}\right| = \frac{16}{3\pi}$.  There are now two
propagating wave fronts superposed.  Each contributes with about
50\% probability, but the additional front propagating to the left has
at each time step a smaller number of flipped spins, which explains the
reduced total spin-flip rate. We thus have the 
astonishing result that a single modified coupling at the end of the chain
can drastically and macroscopically change the dynamics of the entire chain,
even in the limit of arbitrarily long chains. This is a highly unusual
situation, as in most physical systems boundary terms are negligible in the
thermodynamic limit. Exceptions can exist for systems exactly at a phase
transition \cite{DBraun98b}, but this is clearly not the situation for
our spin chains. 

In section
\ref{seq:Comparison-between-Linear-chain-Circular-Chains} we examine
the differences in the 
dynamics for the three different ways of closing the chain in more detail
and elucidate their physical origin. Before doing so,
  we would like to 
point out, however, yet another way of closing the chain, with the rather peculiar property of allowing a dynamics corresponding to
different topologies of the chain depending on the subspace of Hilbert
space considered.

\subsection{Existence of a circular chain dynamically equivalent to
  the linear chain}
Consider the situation where spins $S_2$ to
$S_{N_C-2}$ are coupled through the same nearest-neighbor
interactions as for $H^{(L)}$, but where we have an additional coupling
for spins $S_{N_C-2}$, $S_{N_C-1}$, $S_{N_C}$ and $S_1$ that  
leads to the following action: if spins $S_{N_C-2}$ and $S_1$ are the same and different from $S_{N_C}$, then spin $S_{N_C-1}$ is flipped, otherwise nothing is done. The hamiltonian of such a system is given by
\begin{eqnarray}
H^{(C)}_4&=&\frac{1}{4}\sum_{k=2}^{N_C-2} X_k \left(1_{k-1}\otimes 1_{k+1}- Z_{k-1}\otimes Z_{k+1}\right) \\ \nonumber
&+& \frac{1}{8}
X_{N_C-1}\left(1_{N_C-2}\otimes1_1+Z_{N_C-2}\otimes Z_1\right)\left(1_{N_C}\otimes
  1_1-Z_{N_C}\otimes Z_1\right)\,.
\end{eqnarray}
In this system the states with all spins up or down are still
stationary. One verifies that in the full Hilbert space spanned by the
$2^{N_C}=2^{N_L}$ computational basis states the matrix
representations of $H^{(L)}$ and $H^{(C)}_4$ differ. Nevertheless, it
turns out that the subspaces 
connected to the initial state
$\left|\Lambda_1\right\rangle$ through the dynamics generated by $H^{(L)}$
and $H^{(C)}_4$ are identical for the two hamiltonians.
Furthermore, the matrix representations of $H^{(L)}$
and $H^{(C)}_4$ in these 
parts of Hilbert space connected to 
$\left|\Lambda_1\right\rangle$ are identical, and thus lead to
identical dynamics.  We therefore have the interesting situation that
the different topology of the circular chain manifests itself only in a
certain subspace of Hilbert space, whereas within the subspace
relevant for the spin amplification problem one cannot distinguish
the two hamiltonians through the dynamics which they generate,
whatever the observable.

\color{black}

\section{Comparison of linear chain and circular chains}
\label{seq:Comparison-between-Linear-chain-Circular-Chains}
In the following subsections we compare the time evolution of the
total polarizations as well as the total fidelities with respect to
the two initial states of the first spin for the different
hamiltonians presented above. 
We define the total fidelities $F^{(X)}(t)$ 
($X\in\left\{L,C1,C2,C3\right\}$) for $H^{(L)}$, $H_1^{(C)}$,
$H_2^{(C)}$ and $H_3^{(C)}$, as
\begin{eqnarray}
F^{(X)}(t)&=&F_0^{(X)}(t)+F_1^{(X)}(t)\mbox{ with}\\
F_0^{(X)}(t) &=& \langle
\Lambda_0(t) | \left( \sum_{m=1}^{N_X}
|\uparrow_m\rangle\langle\uparrow_m|\right) |
\Lambda_0(t)\rangle\nonumber\\
F_1^{(X)}(t) &=& \langle \Lambda_1(t) | \left( \sum_{m=1}^{N_X}
|\downarrow_m\rangle\langle\downarrow_m|\right) |
\Lambda_1(t)\rangle\,,
\end{eqnarray}
with  $
|\Lambda_0(0)\rangle=|\Lambda_0\rangle=\left|\uparrow_1\ldots\uparrow_{N_X}\right\rangle$  
and  
$
|\Lambda_1(0)\rangle=|\Lambda_1\rangle=\left|\downarrow_1\uparrow_2\ldots\uparrow_{N_X}\right\rangle$.   
These fidelities can be considered as amplification factors summed
over the two initial basis states. They are a generalization of the usual
fidelity considered in the spin transfer problem, where the fidelity of the
state of the final spin with respect to the pure initial states of the first
spin are considered \cite{Bose03}. As $H^{(L)}$, $H_1^{(C)}$,  and
$H_2^{(C)}$ conserve the basis state $|\Lambda_0\rangle$ we will see
that for them the total fidelity is directly related to the average
total polarization.  However, $H_3^{(C)}$ does not conserve
$|\Lambda_0\rangle$.   In
this situation, the total fidelities are useful for comparing
the overall performance of the different devices as spin amplifier.
Note that all fidelities have initial value $N_L+1$ ($N_C+1$), and are
bounded from above by $2N_L$ ($2N_C$) for linear (circular) chains.

\color{black}
\subsection{Comparison of $H^{(L)}$ and $H_3^{(C)}$}
\begin{figure}[ht]
\includegraphics[width=9cm]{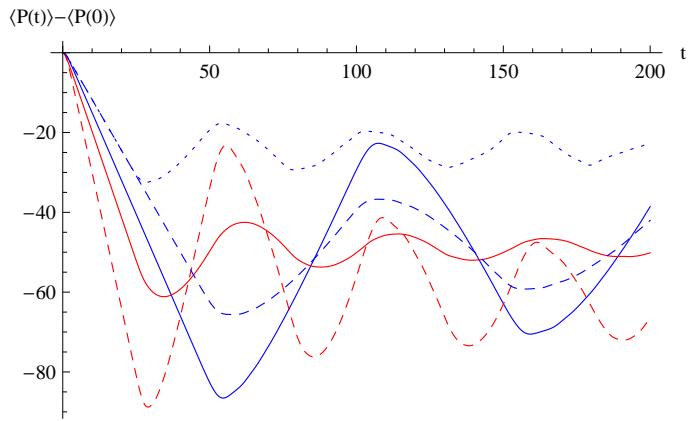}  
\caption{(Color online) Comparison of mean polarizations $\left \langle
  P(t)\rangle -\langle P(0)\right \rangle $ for linear chain
  $H^{(L)}$ (continuous blue line), circular chain $H^{(C)}_3$ with $N_C=N_L$ (dashed
  blue line), circular chain $H^{(C)}_3$ with $N_C=(N_L+1)/2$ (dotted
  blue line), and circular chains $H_1^{(C)}$ (dashed red line), $H_2^{(C)}$ (continuous red line) with $N_C=N_L=51$}
\label{fig:comparison-relative-mean-polarizations}
\end{figure}

In Fig.\ref{fig:comparison-relative-mean-polarizations} we compare the time
evolution of the mean polarizations for the linear chain $H^{(L)}$ studied
in section \ref{seq:Linear-chain} with basis of size $N_L-1$, and the
circular chain $H_3^{(C)}$ studied in section
\ref{seq:Closing-by-particular-coupling} with basis of size $2(N_C-1)$. We
first consider the case $N_C=N_L$. We observe that the mean polarizations
for the two types of chains oscillate with approximately the same
frequency. For $t\apprle N_L$ the mean polarizations decrease linearly, but
with a difference in the slopes of $25\%$ as predicted by
Eq.(\ref{eq:Linear-chain-WKB-Polarization-Linear-term}) and
Eq.(\ref{eq:Circular-chain-WKB-polarization-Linear-term}). The question of
the physical origin of the different behaviors arises. The simulations shown
in Fig.\ref{fig:comparison-relative-mean-polarizations} were done for $N_L =
N_C=51$. 

Adding an additional local term to the hamiltonian for closing the
chain would be expected to lead at most to $1/N_L \sim 2\%$
effect. Moreover, our analytical results show that the $25\%$ effect
persists for $N_L \rightarrow \infty$. 
A first guess about the origin of the different behavior might be the
dimension of the involved basis sets. Since the dimension of the basis
 for $H^{(L)}$  
($\left\{\left|\Lambda_k\right\rangle\right\}$ with $k\in[1, N_L-1]$, seen
in section \ref{seq:Linear-chain-Description-of-system}) is half of
the one of $H_3^{(C)}$ ($\left\{\left|\Lambda_k\right\rangle\right\}$ plus
the corresponding mirror states
$\left\{\left|\overline{\Lambda_k}\right\rangle\right\}$ with $k\in[1,
  N_C-1]$, as seen in section \ref{seq:Closing-by-particular-coupling}), we
might think that this is a reason for the observed difference. In order to check
this hypothesis, we recalculated the polarization for the 
circular chain for $N_C = (N_L+1)/2$ in which case the two bases have the
same dimension. In order to compare the polarization of chains of different
length, we subtract the initial polarization, i.e.~we consider the change
of polarization with respect to the initial state. As
Fig.\ref{fig:comparison-relative-mean-polarizations} shows, we observe the
same difference between the slopes of the two mean polarizations as function
of time (for $t\apprle N_L/2$) as before. Again, this is confirmed by the
leading terms in Eq.(\ref{eq:Linear-chain-WKB-Polarization-Linear-term}) and
Eq.(\ref{eq:Circular-chain-WKB-polarization-Linear-term}), which are
independent of $N_L$, $N_C$. This shows that the difference of basis sizes
is not responsible for the different behavior of the two systems.
The matrix representations of the two
hamiltonians
$H^{(L)}$  and
$H_3^{(C)}$
, Eqs.(\ref{eq:Linear-chain-Matrix-representation-of-hamiltonian}) and
(\ref{eq:Circular-chain-Closing-by-particular-coupling-Matrix-representation-of-hamiltonian}),   
differ only by the two off-diagonal matrix elements $H_{1,N_L-1}=
H_{N_L-1,1}$, which equal $1$ for the circular chain, but $0$ for the linear
chain. Thus, one would expect at worst a correction of ${\cal O}(N_L)$ to the
eigenvalues and eigenstates of $H_3^{(C)}$. This is confirmed by numerical
diagonalization, which indicates that the average absolute difference
between corresponding matrix elements of the two propagators decays even
more rapidly, roughly as $1/N_L^2$.  Clearly, the slightly different matrix
representations cannot explain the observed macroscopic differences either.
Fig.\ref{fig:comparison-relative-mean-polarizations} also shows that the
mean polarization of the circular chain oscillates approximately twice as
fast for $N_C=(N_L+1)/2$ compared to $N_C=N_L$. This makes sense, as the
oscillation period is set by the time it takes for the spin waves to reach
the end of the chain. Also, for $N_C=(N_L+1)/2$, the amplitude of the signal
is roughly half the one for $N_C=N_L$ since the number of spins of the
circular chain has been divided by 2 (up to one spin). 

All of this shows
that the important difference in the two models lies in the physical
structure of the basis functions. The single additional coupling in
$H_3^{(C)}$ allows access to a new part of Hilbert space (i.e.~the
additional basis states $\left| \overline{\Lambda_k}\right\rangle$). Since
the $\left|\overline{\Lambda_k}\right\rangle$ have opposite polarization
compared to $\left|\Lambda_k\right\rangle$, their admixture reduces the
total polarization signal compared to the linear chain. Thus, a kind of
bifurcation takes place in Hilbert space, depending on the presence of the
additional coupling, and the difference in the physical properties of the
additionally accessible basis states leads to the macroscopically different
behavior.

\begin{figure}[ht]
\includegraphics[width=9cm]{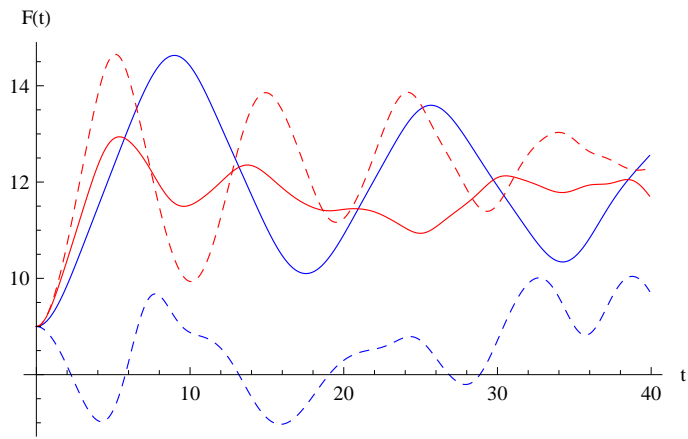}  
\caption{(Color online) Comparison of total fidelities $F(t)$ for linear chain
  $H^{(L)}$ (continuous blue line), circular chain $H^{(C)}_3$ (dashed
  blue line), and circular chains $H_1^{(C)}$ (dashed red line), $H_2^{(C)}$ (continuous red line), with $N_C=N_L=8$}
\label{fig:comparison-total-fidelities-of_polarization}
\end{figure}

In Fig.\ref{fig:comparison-total-fidelities-of_polarization} we
show the total fidelities. For $H^{(L)}$,
$F^{(L)}(t)$ behaves 
like the inverted polarization $\langle
P^{(L)}(t)\rangle$, scaled  by a factor
$1/2$ and shifted by a constant. Indeed we have $
F^{(L)}(t)=F_0^{(L)}(t)+F_1^{(L)}(t)$ with
$F_0^{(L)}(t)=\frac{N_L}{2}+\frac{1}{2}\langle\Lambda_0(t)|
P^{(L)}| \Lambda_0(t) \rangle$ and
$F_1^{(L)}(t)=\frac{N_L}{2}-\frac{1}{2}\langle
\Lambda_1(t)|P^{(L)}| \Lambda_1(t) \rangle$. But 
since $\left| \Lambda_0 \right\rangle$ is stationary $F_0^{(L)}(t)$
reduces to $N_L$, and so 
\begin{eqnarray}
\label{eq:Fidelity-of-polarization-for-linear-chain}
F^{(L)}(t)= \frac{3}{2}N_L -\frac{1}{2}\langle P^{(L)}(t)\rangle
\end{eqnarray}
For $H_3^{(C)}$ we observe that the total fidelity 
immediately deteriorates (whereas $F^{(L)}(t)$ increases initially),
and remains much smaller than  $F^{(L)}(t)$. This behavior
results from $F_0^{(C3)}(t)$ which is not constant, as
$\left|\Lambda_0\right\rangle$ is no longer a stationary 
state.  The dynamics generated by $H_3^{(C)}$ starting from 
$\left|\Lambda_0\right\rangle$ evolves apparently in a subspace of
dimension $2^{N_C-2}$ (verified numerically up to $N_C=9$). In order
to compare the total fidelities for the different chains on equal
footing, all fidelities have been plotted in
Fig.\ref{fig:comparison-total-fidelities-of_polarization} for chains
of 8 spins. 
Not shown in Fig.\ref{fig:comparison-total-fidelities-of_polarization}
is that
if we wait a sufficiently long time, the total 
fidelity $F^{(C3)}(t)$ can become large again for particular values of
$t$. As an example, for $N_C=8$ we obtain $F^{(C3)}(290)\approx
13.09$, close to the maximal possible value 16 for this
chain. However, comparisons of the fidelity should be made for a given
fixed time interval, as for sufficiently long times a
spin-configuration arbitrarily close to a given one can be found \cite{Bose03}.

\color{black}
\subsection{Comparison of $H^{(L)}$, $H_1^{(C)}$, and $H_2^{(C)}$ }
\begin{figure}[ht]
\includegraphics[width=9cm]{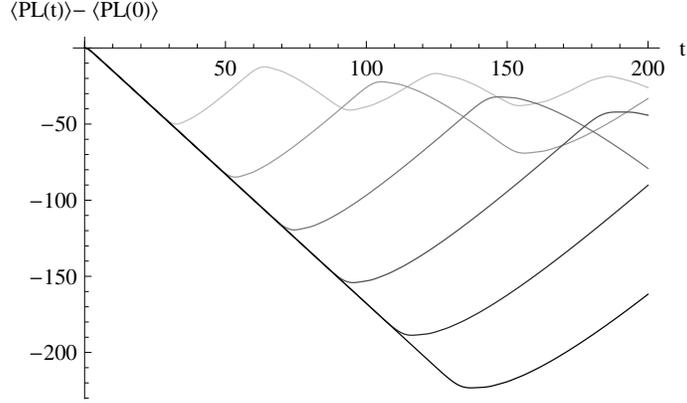} 
\caption{(Color online) Mean polarizations  $\langle
  P^{(L)}(t)\rangle -\langle P^{(L)}(0)\rangle $ for
  linear chains $H^{(L)}$ for different chain sizes (from $N=30$ (top curve) to
  $N_L=130$ (bottom curve) in steps of $20$)} 
\label{fig:PL_Relative}
\end{figure}

\begin{figure}[ht]
\includegraphics[width=9cm]{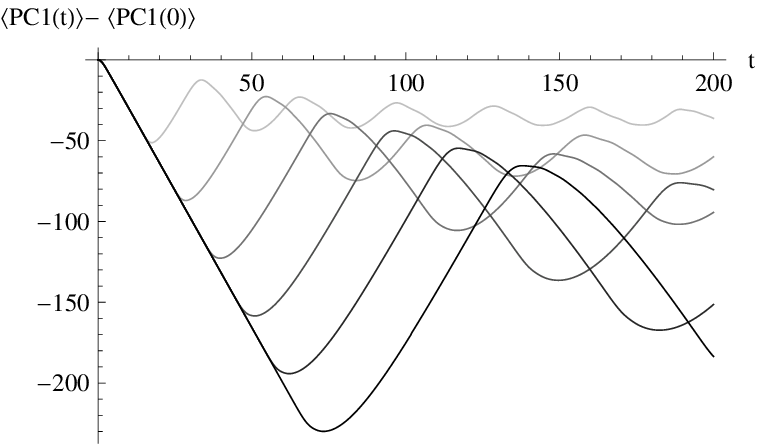} 
\caption{(Color online) Mean polarizations  $\langle
  P^{(C)}_1(t)\rangle -\langle P^{(C)}_1(0)\rangle $ for
  circular chains $H^{(C)}_1$ for different chain sizes (from $N=30$ (top
  curve) to
  $N_L=130$ (bottom curve) in steps of $20$)} 
\label{fig:PC1_Relative}
\end{figure}

\begin{figure}[ht]
\includegraphics[width=9cm]{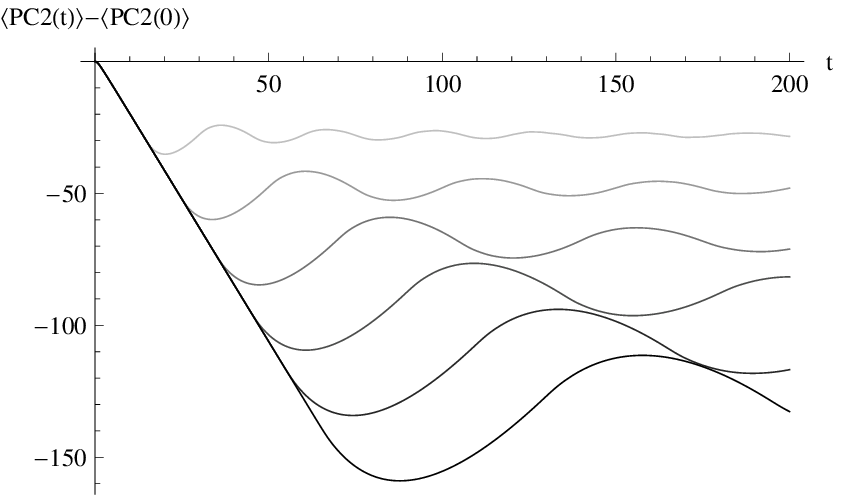} 
\caption{(Color online) Mean polarizations $\langle
  P^{(C)}_2(t)\rangle -\langle P^{(C)}_2(0) \rangle $ for
  circular chains $H^{(C)}_2$ for different chain sizes (from $N=30$ (top
  curve) to
  $N_L=130$ (bottom curve) in steps of $20$)} 
\label{fig:PC2_Relative}
\end{figure}

\begin{figure}[ht]
\includegraphics[width=7cm]{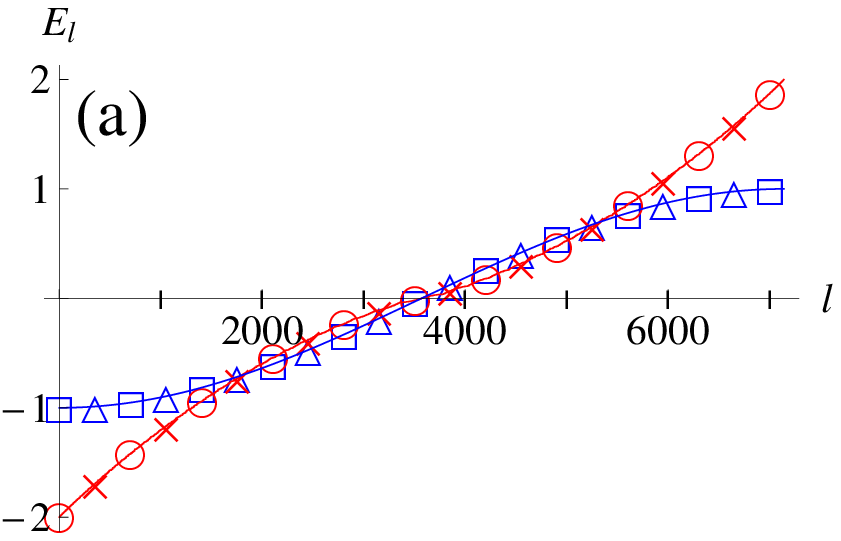}
\includegraphics[width=7cm]{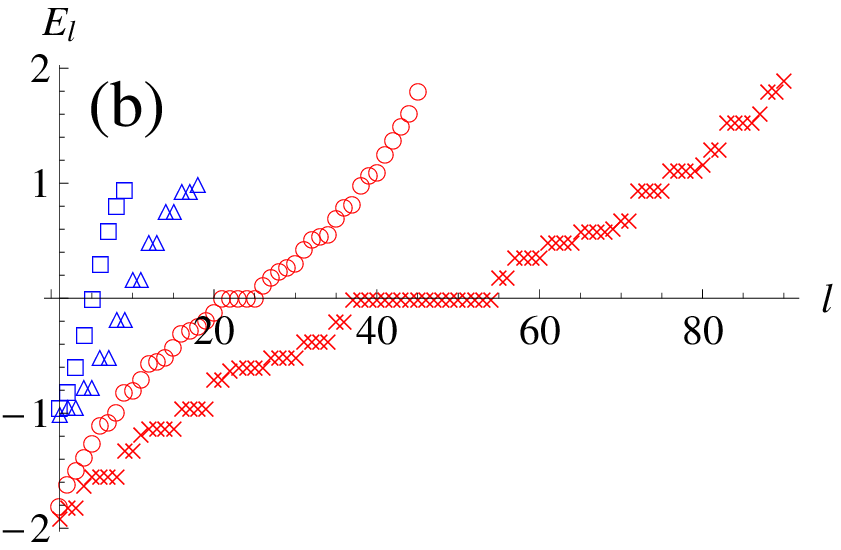} 
\caption{(Color online) (a) Spectra for $H^{(L)}$   (blue squares, $N_L=7141$), $H_1^{(C)}$ (continuous red line with crosses, $N_C=120$),
$H_2^{(C)}$ (red circles, $N_C=80$), $H_3^{(C)}$ (blue triangles,
  $N_C=3571$). The cosine dispersion law of
  Eq.(\ref{eq:Linear_Chain_Eigenvalues}) is plotted as continuous blue
  line. Only a subset of the ordered set of all eigenvalues $E_l$ is plotted
  as function of $l$  in order to show the 
  dispersion relation. (b) Spectra for  $H^{(L)}$, $H_1^{(C)}$, $H_2^{(C)}$
  and $H_3^{(C)}$ for $N_L=N_C=10$. Same symbols as in (a).}    
\label{fig:Spectra}
\end{figure}

In Fig.\ref{fig:PL_Relative}, Fig.\ref{fig:PC1_Relative}, and
Fig.\ref{fig:PC2_Relative},  we plot the  behavior of mean polarizations
$\langle P^{(L)}\rangle $, $\langle P^{(C)}_1
\rangle$ and $\langle P^{(C)}_2
\rangle$ for chains of different lengths. For $H_1^{(C)}$ and
$H_2^{(C)}$, the mean polarizations were calculated by numerical
diagonalization since generalizing the semiclassical approach to these
hamiltonians would be rather cumbersome. For $\langle
P^{(L)}\rangle$, $\langle P^{(C)}_1\rangle$ and $\langle P^{(C)}_2\rangle$ we observe that
the slopes of the curves for sufficiently small times, where the mean
polarizations 
behave linearly, are independent of the size of the chains.  They equal
$\left|v^{(L)}\right|=\frac{16 }{3\pi}\approx 1.69$ (obtained analytically, see above)
in the case of $H^{(L)}$,
$\left|v^{(C)}_1\right|\approx 3.39$ and $\left|v^{(C)}_2\right|\approx 2.16$ (obtained
numerically) in the case of $H^{(C)}_1$ and $H^{(C)}_2$. Therefore, the polarizations of 
circular chains  $H^{(C)}_1$ and  $H^{(C)}_2$ reach their first minimum respectively about $2$ and $1.28$ times faster
than the linear ones.  
Since $\left|\Lambda_0\right\rangle$
remains stationary for $H_1^{(C)}$ and $H_2^{(C)}$, the  total
fidelities $F^{(C1)}$  and $F^{(C2)}$ follow the same law as 
Eq.(\ref{eq:Fidelity-of-polarization-for-linear-chain}) with
corresponding mean polarizations (see
Fig.\ref{fig:comparison-total-fidelities-of_polarization}).
\color{black}
The origins of the different behaviors compared to $H^{(L)}$ are more
complex here.  First of all, 
the dimensions of the set of basis states connected by $H_1^{(C)}$,
$H_2^{(C)}$ and $H^{(C)}_3$ already differ in their scaling with the length
of the chains (scaling as ${\cal O}(N_C^2)$ for  $H_1^{(C)}$,
$H_2^{(C)}$ and as ${\cal O}(N_C)$ for $H^{(C)}_3$). Secondly, the matrix
representations are substantially different, and in fact nonequivalent, such
that the spectra of the hamiltonians in the accessible Hilbert spaces are
different. This can be observed in Fig.\ref{fig:Spectra} where we
compare the spectra for the  
different hamiltonians.  The size of the different chains have been
chosen  
such that the basis of $H^{(L)}$, $H_1^{(C)}$, $H_2^{(C)}$ and $H_3^{(C)}$
are of same dimension ($7140$ basis states with $N_L=7141$, $N_{C1}=120$, $N_{C2}=80$, $N_{C3}=3571$, with $N_{C1}$, $N_{C2}$, $N_{C3}$, the sizes of the circular chains corresponding to $H_1^{(C)}$, $H_2^{(C)}$ and $H_3^{(C)}$). We can observe that the spectra
for  $H^{(L)}$ and $H_3^{(C)}$ roughly coincide, as do the spectra for
$H_1^{(C)}$ and $H_2^{(C)}$. For $H^{(L)}$ and $H_3^{(C)}$ we have seen in
subsections \ref{seq:Linear-chain-Description-of-system} and
\ref{seq:Closing-by-particular-coupling} that the energy levels are given by
a 1D tight-binding model with nearest-neighbors hopping,
and thus follow the cosine laws in Eq.(\ref{eq:Linear_Chain_Eigenvalues})
and Eq.(\ref{eq:HC3_Eigensystem}).  For $H_1^{(C)}$, 
$H_2^{(C)}$ we have in the limit of large $N_C$ for most states essentially
a 2D tight-binding model: All states inside the pyramid Eq.(\ref{eq:Triangle-of-States}) are
coupled  
to four nearest neighbors, and the same is true for the states on the
vertices of all polygons in Fig.\ref{fig:states_connections_for_H2} for $H_2^{(C)}$, with the exception
of the innermost and the outermost polygons.  However, the spectra differ
substantially here from the usual $\cos(k_x)+\cos(k_y)$ spectra for a
tight-binding model on a square lattice due to the
unusual geometry. For example, for $N_C=5$, the lattice corresponding to
$H_2^{(C)}$ has the dihedral symmetry of a pentagon ($C_{5v}$, see
Fig.~\ref{fig:states_connections_for_H2}). This symmetry leads to rather
large degeneracies, in particular in the center of the spectrum, and to
finite slopes of the dispersion relation at the edges of the spectrum, which
are particularly visible for relatively small chain lengths (see
Fig.~\ref{fig:Spectra}). Finally, also the relevant basis states are once more
different from those of $H^{(L)}$.
\section{Conclusion}
In this article, we have shown how the way of closing a linear chain into a
circular chain can lead to significantly different dynamical behavior. A
single additional coupling may open access to additional parts of Hilbert
space which may differ in their dimensions (and even the scaling of the
dimensions with the number of spins), the physical properties of the
basis functions, or give rise to different, nonequivalent matrix
representations of the hamiltonian.   This can modify the time-dependent 
polarizations on a macroscopical level, even though the hamiltonians only
differ locally by one spin and its coupling to the rest of the chain. For
example, the rates of change of polarization in the linear regime for
$H_3^{(C)}$ and $H^{(L)}$ (see
Eqs.(\ref{eq:Circular-chain-WKB-polarization-Linear-term}) and
(\ref{eq:Linear-chain-WKB-Polarization-Linear-term})) differ by about
$25\%$, even  
for arbitrarily long chains. In this case, the number of basis states of
$H_3^{(C)}$ is twice as large as the number of basis states of $H^{(L)}$ but
remains of the same order ${\cal O}(N_L)$. When adjusted to the same dimension of
the basis by changing the length of the chain, almost the
same representation of the hamiltonian results, but the physical properties of
the basis functions differ strongly. 
The rates of polarization change for $H_1^{(C)}$ and $H_2^{(C)}$ differ even
from the one of $H^{(L)}$, by roughly a factor $2$ and $1.28$,
respectively.  Here, the number of basis states connected by $H_1^{(C)}$ or
$H_2^{(C)}$  scales as 
${\cal O}(N_C^2)$. The matrix
representations of the hamiltonians have a substantially different
structure, and also the physical properties of the basis states differ. 
It remains to be seen whether applications can be found which exploit the
high sensitivity of the time dependent polarizations to the additional local
couplings used to change the boundary conditions.  
\acknowledgments{We thank CALMIP (Toulouse) and IDRIS (Orsay) for
the use of their computers. B.~Roubert is supported by a grant from the
DGA, with M.~Jacques Blanc--Talon as scientific liaison officer. P.~Braun is grateful to the Sonderforschungsbereich TR 12 of the 
Deutsche Forschungsgemeinschaft and to the GDRI-471.}

\bibliography{../../mybibs_bt}
\end{document}